\documentclass[aps,prb,twocolumn,10pt]{revtex4-1}

\usepackage{graphicx}
\usepackage{amsmath}
\usepackage{soul}

\begin{document}
\title{Maxon and roton measurements in nanoconfined $^4$He}

\author{M. S. Bryan}
\email[]{msbryan@indiana.edu}
\author{P. E. Sokol}
\affiliation{Department of Physics, Indiana University, Bloomington, IN 47408, USA}
\date{\today}

\begin{abstract}
We investigate the behavior of the collective excitations of adsorbed $^4$He in an ordered hexagonal mesopore, examining the crossover from a thin film to a confined fluid. Here we present the inelastic scattering results as a function of filling at constant temperature. We find a monotonic transition of the maxon excitation as a function of filling. This has been interpreted as corresponding to an increasing density of the adsorbed helium, which approaches the bulk value as filling increases. The roton minimum exhibits a more complicated behavior that does not monotonically approach bulk values as filling increases. The full pore scattering resembles the bulk liquid accompanied by a layer mode. The maxon and roton scattering, taken together, at intermediate fillings does not correspond to a single bulk liquid dispersion at negative, low, or high pressure.   
\end{abstract}
\maketitle
\section{Introduction}

The structure and density of liquids in confinement is of great interest in a wide range of phenomena, ranging from the liquid-liquid phase transition in water\cite{liu2007observation,mallamace2007evidence}, molecular layering in thin films\cite{yu1999observation,huisman1997layering}, supercritical fluids in confinement\cite{melnichenko2009characterization,rother2007microstructural}, and confined glass formers\cite{morineau2002finite,scheidler2000relaxation}. In bulk liquid $^4$He, the structure and density are closely related to the excitation spectrum\cite{feynman1954atomic}, the phonon-maxon-roton spectrum, which has been extensively measured\cite{cowley1971inelastic}. Liquid helium, a simple atomic liquid with well-known interactions, is a model system to explore the effects of confinement on liquid structure using the energies of the excitations as a spectroscopic probe of the local fluid density in a restricted geometry.

Previous studies of confined helium in large pore systems (diameter $>$ 7 nm) such as aerogel\cite{azuah1999excitations,faak2000phonons,plantevin1998elementary,coddens1995time}, xerogel\cite{anderson2002dynamics}, and Vycor\cite{glyde2000dynamics,dimeo1998localized,plantevin2001excitations} have observed excitation spectra in agreement with the bulk liquid. However, for small pores a recent neutron spectroscopy study concluded that liquid in partially filled pores corresponded to bulk liquid helium at negative absolute pressures \cite{albergamo2004phonon,albergamo2005elementary}.  Albergamo \textit{et al.} observed that the energy of the maxon is shifted downward relative to the corresponding bulk value, and as the pore space is filled with further liquid, the energy of the maxon approaches the bulk value from below.  By extrapolating the behavior of the bulk liquid to the nanoconfined case, they reported negative pressures down to -5.5 bar can be realized in pores of diameter 47 $\textrm{\AA}$.  Negative pressure states are metastable against cavitation, and have been measured extensively in $^4$He\cite{imre2002liquids}. A realization of negative pressure in porous media would constitute a significant amount of negative pressure liquid available for study, with much longer lifetimes, than in typical ultrasonic studies. Unfortunately, their measurements did not extend to the roton region of the spectrum and so the filling dependence of the roton was not reported.

\begin{figure}
	\includegraphics[width=\linewidth]{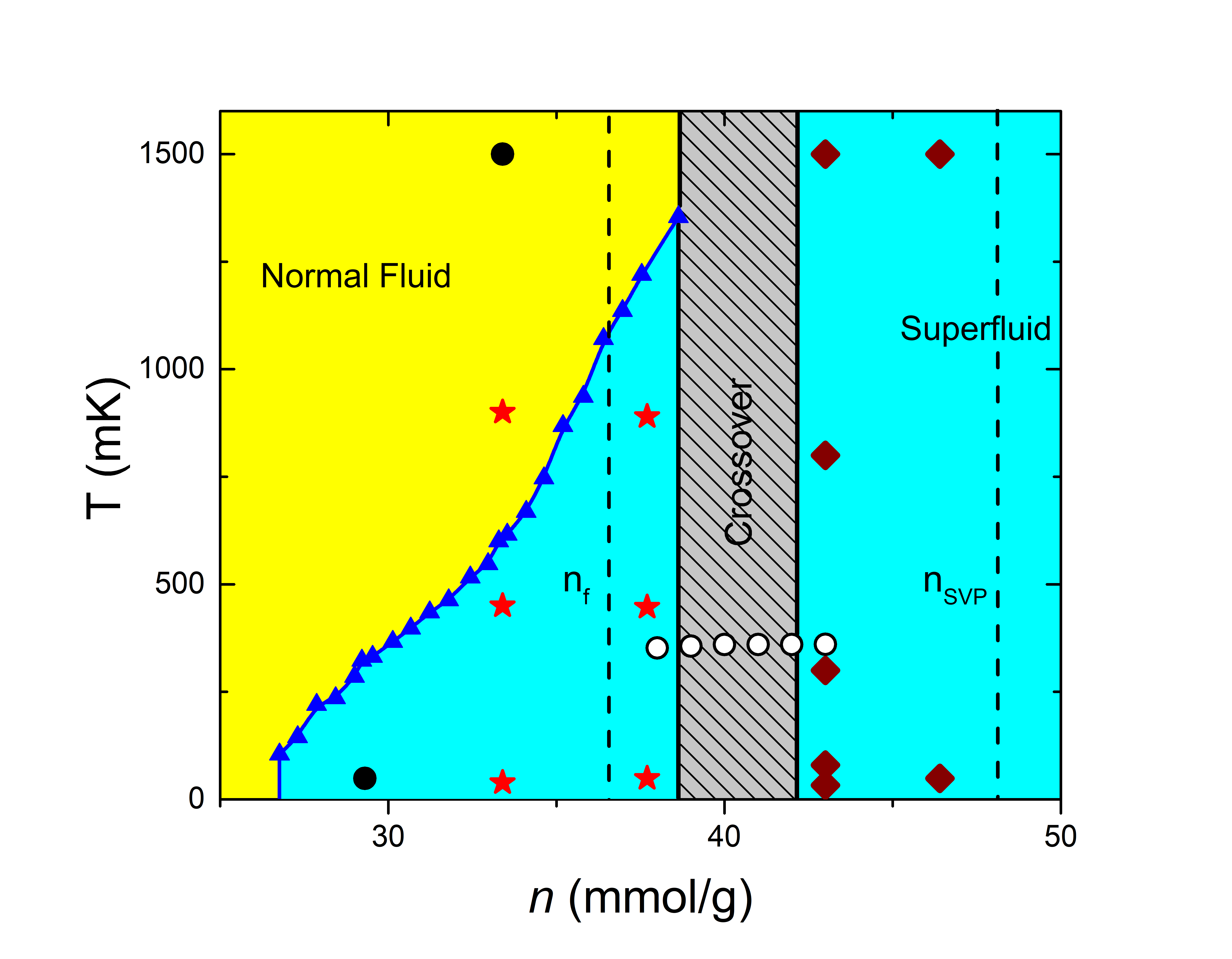}
	\centering
	\caption{Previously reported phase diagram\cite{prisk2013phases}. The superfluid transition line (blue triangles) is taken from previous torsional oscillator measurements of Ikegami \textit{et al}\cite{ikegami2007superfluidity2}. Inelastic neutron scattering with no well defined excitations are indicated by black circles. Scattering corresponding to a low density thin film (red stars) are shown  to the left of the crossover region. Bulk-like scattering, in agreement with the bulk liquid scattering under vapor pressure (dark red diamonds), is shown  to the right of the crossover region. Inelastic neutron scattering measurements reported in this paper are shown as white circles, moving horizontally through the shaded gray crossover region. }
	\label{fig:PhaseDiagram}
	\end{figure}

Recently, Prisk \textit{et al.} measured the excitation spectrum as a function of temperature and filling in another small pore system\cite{prisk2013phases}.  They identified two regions - a low density phase and a high density phase, corresponding to a thin film and bulk-like excitation spectrum.  Within these regions the maxon and roton energy was independent of filling.  They interpreted this in terms of two different phases with a crossover regime between them as shown in Fig. \ref{fig:PhaseDiagram}. The low and high density phases would correspond to pressures of -7.5 and 0 bar, respectively, based on the maxon energy. However, this study did not address the intermediate filling range studied by Albergamo \textit{et al}.

In this paper we present inelastic neutron scattering measurements of the phonon-roton spectrum of superfluid helium within an ordered mesoporous silica known as Folded Sheet Material (FSM-16). We report simultaneous measurements of the maxon and roton energies as a function of pore filling in the crossover region, between the thin film and bulk-like regions.  Following Albergamo \textit{et al.}, we have attempted to interpret the pore filling dependence of these excitations by extrapolating the density dependence of the bulk fluid to the nanoconfined case.  However, we find that a self-consistent interpretation of the scattering data cannot be obtained when the extrapolation is simultaneously applied to the maxon and the roton.  We therefore infer that the structure of superfluid helium, when partially filling pore spaces only a few nanometers in diameter, is not consistent with viewing the confined liquid as a bulk fluid under negative pressure. 

\section{Experimental Approach}

FSM-16, a monodisperse porous silica with hexagonal pores arranged on a triangular lattice, provided the confining media for $^4$He in these measurements. A detailed description of the synthesis process of FSM-16 has been reported elsewhere\cite{inagaki1996syntheses}. This sample was used in a previous study\cite{prisk2013phases} and has been characterized via N$_{2}$ and $^4$He adsorption isotherms. The nitrogen isotherm was type IV and yielded a BET surface area of 1015 m$^2$/g and a relatively narrow pore-size distribution centered at 27 $\textrm{\AA}$ with a full width half-max of approximately 2 $\textrm{\AA}$. The helium isotherm yielded a monolayer coverage of 21.9 mmol/g and a full pore filling at approximately 43 mmol/g. The structure of dry FSM-16 was characterized by x-ray diffraction which was consistent with a 2D triangular lattice (space group symmetry \textit{p6mm}) with a lattice constant of $\textit{a} \approx$ 45 $\textrm{\AA}$. The FSM-16 was heated to 100 $^{\circ}$C for 24 hours and pumped to P $<$ $10^{-2}$ torr to remove surface contamination, primarily water vapor, prior to the measurements.

Inelastic neutron scattering was performed at the Cold Neutron Chopper Spectrometer (CNCS) located at the Spallation Neutron Source\cite{ehlers2011new}. CNCS is a direct geometry time of flight spectrometer that views a cold moderator. An incident energy $E_i$ = 3.65 meV was selected to avoid Bragg scattering from the Al sample cell.  This incident energy yielded a dynamic window of approximately 0 $<$ Q $<$ 2.2 $\textrm{\AA}^{-1}$ momentum transfer and 0 $<$ E $<$ 3 meV energy transfer. The resulting elastic line resolution was 81 $\mu$eV. Standard data reduction routines were used to reduce the raw time-of-flight data to S(Q,E)\cite{arnold2014mantid}, which was subsequently analyzed using the DAVE software package\cite{azuah2009dave}.

The FSM-16 sample was contained in a cylindrical Aluminum sample can with 5.71 cm height and 2 cm diameter.  A continuous flow helium cryostat with an Oxford HelioxVT $^3$He insert was used to cool the sample. Gas was loaded into the cell via a small capillary using standard volumetric techniques. A small heater was placed on the capillary near the sample cell to eliminate superfluid in the fill line which would give an undesirable heat load on the cell.  

\section{Results}
\begin{figure*}
	\includegraphics[width=\linewidth]{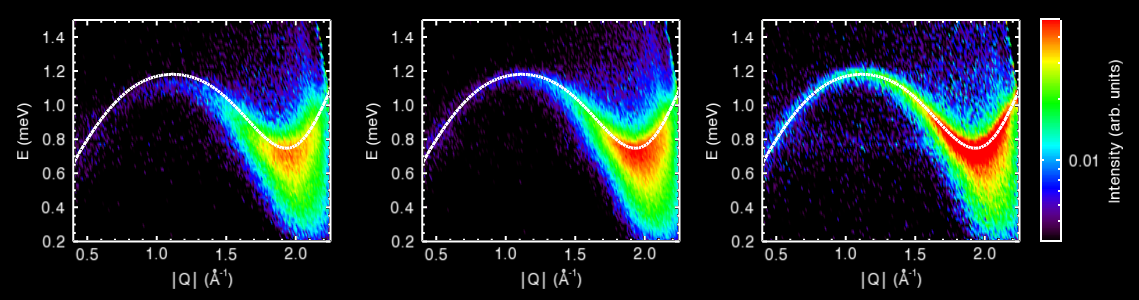}
	\caption{Dynamic structure factor S(Q, E) for 38.0, 40.0, and 43.0 mmol/g with monolayer background subtracted. All plots share a logarithmic intensity scale, with the bulk phonon-roton spectrum shown as a white dotted line. }
	\label{fig:Contour}
	\end{figure*}
    
Neutron scattering measurements were carried out at T = 350 $\pm$ 8 mK with gas loadings from 38.0 mmol/g to 43.0 mmol/g, in 1.0 mmol/g increments.  A measurement at monolayer filling of $n_1$ = 21.9 mmol/g, as determined by helium adsorbtion isotherms, was also performed.  Previous neutron scattering studies have indicated that this first adsorbed layer is solid\cite{prisk2013phases,bossy2012phonon,albergamo2004excitations,albergamo2003elementary} and torsional oscillator measurements indicate that the onset of superfluidity occurs only after the completion of the first solid layer\cite{ikegami2007superfluidity2}. The measured monolayer scattering was broad and featureless, consistent with an inert adsorbed solid.   

The dynamic structure factor S(Q,E) for three fillings are shown in Fig. \ref{fig:Contour}.  The bulk liquid phonon-roton curve is also shown for comparison, as a white dotted line. The monolayer filling measurement was used as background and subtracted from the data. This includes scattering from the porous material and sample cell, as well as the broad, featureless scattering that results from monolayer $^4$He itself. At all fillings studied the excitation spectrum exhibits the characteristic phonon-maxon-roton behavior of the superfluid phase, accompanied by additional scattering known as the layer mode\cite{dimeo1998localized,plantevin2001excitations,faak2000phonons,thomlinson1980excitations}. The layer mode is a relatively broad excitation that spans the roton region in $Q$, with energies lower than the bulk roton energy. The energies of the bulk-like excitations, even far away from the roton minimum and layer mode, are different than the bulk values at low fillings. As the filling is increased the overall intensity of the scattering increases and the observed excitation energies move closer to the bulk excitation energies.  Finally, at the highest filling studied the excitation energies across the entire dynamic window are in good agreement with the bulk values.

The highest filling has a weak feature at roton energies (approximately 0.75 meV) across all $Q$, particularly near the maxon $Q$, known as the ghost roton. The ghost roton is produced from multiple scattering that includes one roton and one elastic scattering event, from the FSM-16 or adsorbed $^4$He. This $Q$-independent feature was fit to a common energy and width across all $Q$, then subtracted from the data. The ghost roton is approximately 100 times less intense than the scattering near the bulk roton, and is not noticeable in that region of intense scattering. In the phonon region of the dispersion, the $^4$He scattering is much weaker and as a result, a direct observation of the speed of sound is not possible with this data set. 

\begin{figure}
	\includegraphics[width=\linewidth]{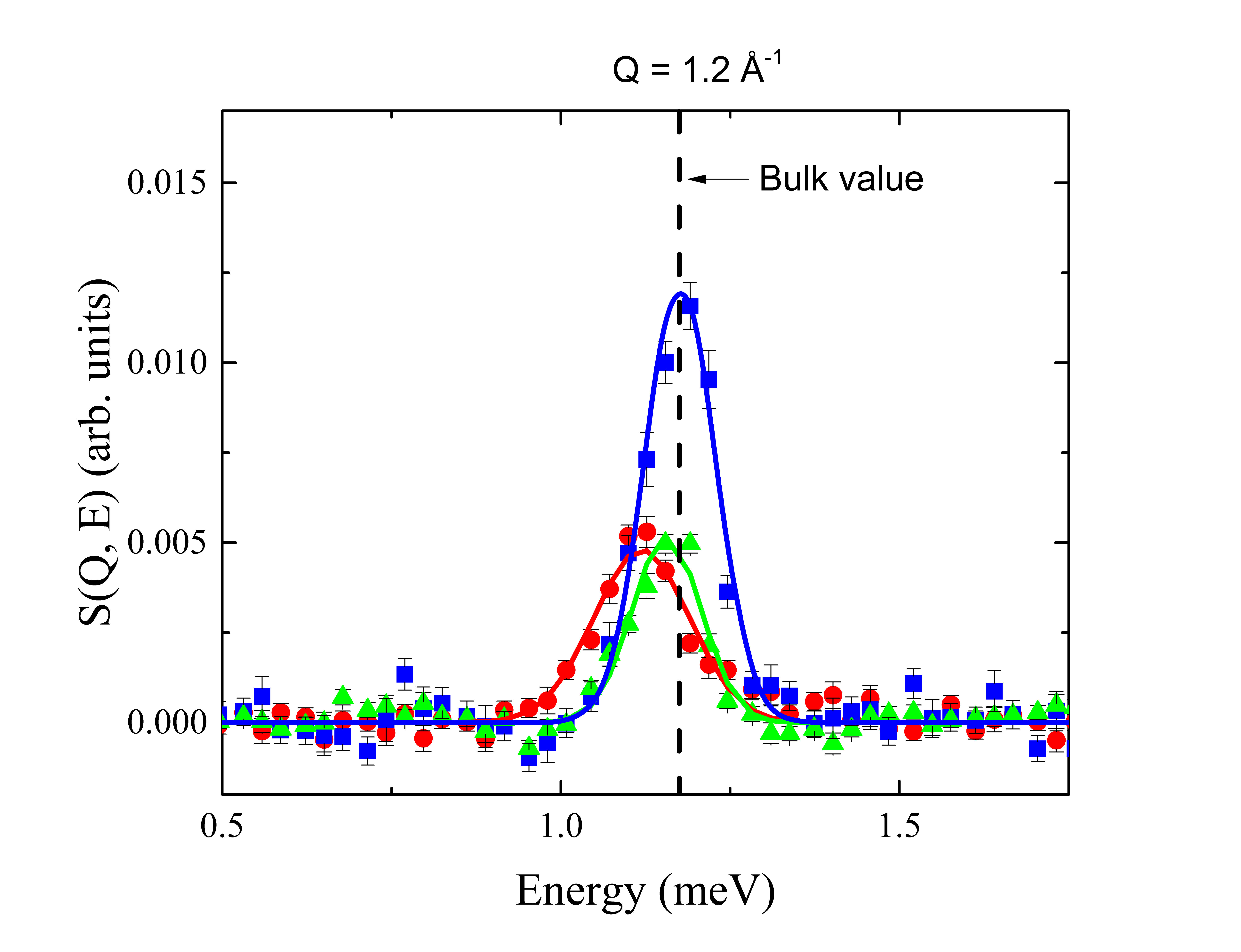}
	\caption{A cut along energy of the maxon, Q = 1.2 $\textrm{\AA}^{-1}$. Data for 38.0, 40.0, and 43.0 mmol/g fillings are shown in red circles, green triangles, and blue squares respectively. Corresponding fits are shown as a colored line. The peak center of the bulk liquid scattering is shown as a vertical dashed line. Background and the ghost-roton have been subtracted, leaving only the maxon peak scattering and corresponding fit.}
	\label{fig:MaxonFits}
	\end{figure}

Constant Q cuts through the scattering at the maxon ($Q$=1.2 $\textrm{\AA}^{-1}$) are shown in Fig. \ref{fig:MaxonFits} for three fillings, with the ghost roton and background subtracted. Points far away from the maxon peak are generally centered near zero, demonstrating a reasonable background subtraction. At the lowest filling studied (38.0 mmol/g) the maxon energy is significantly below the bulk value. As filling increases, the center of the peak moves to higher energies and approaches the bulk liquid value. The maxon intensity slightly decreases initially from 38.0 to 40.0 mmol/g, then increases dramatically from 40.0 to 43.0 mmol/g. 

\begin{figure}
	\includegraphics[width=\linewidth]{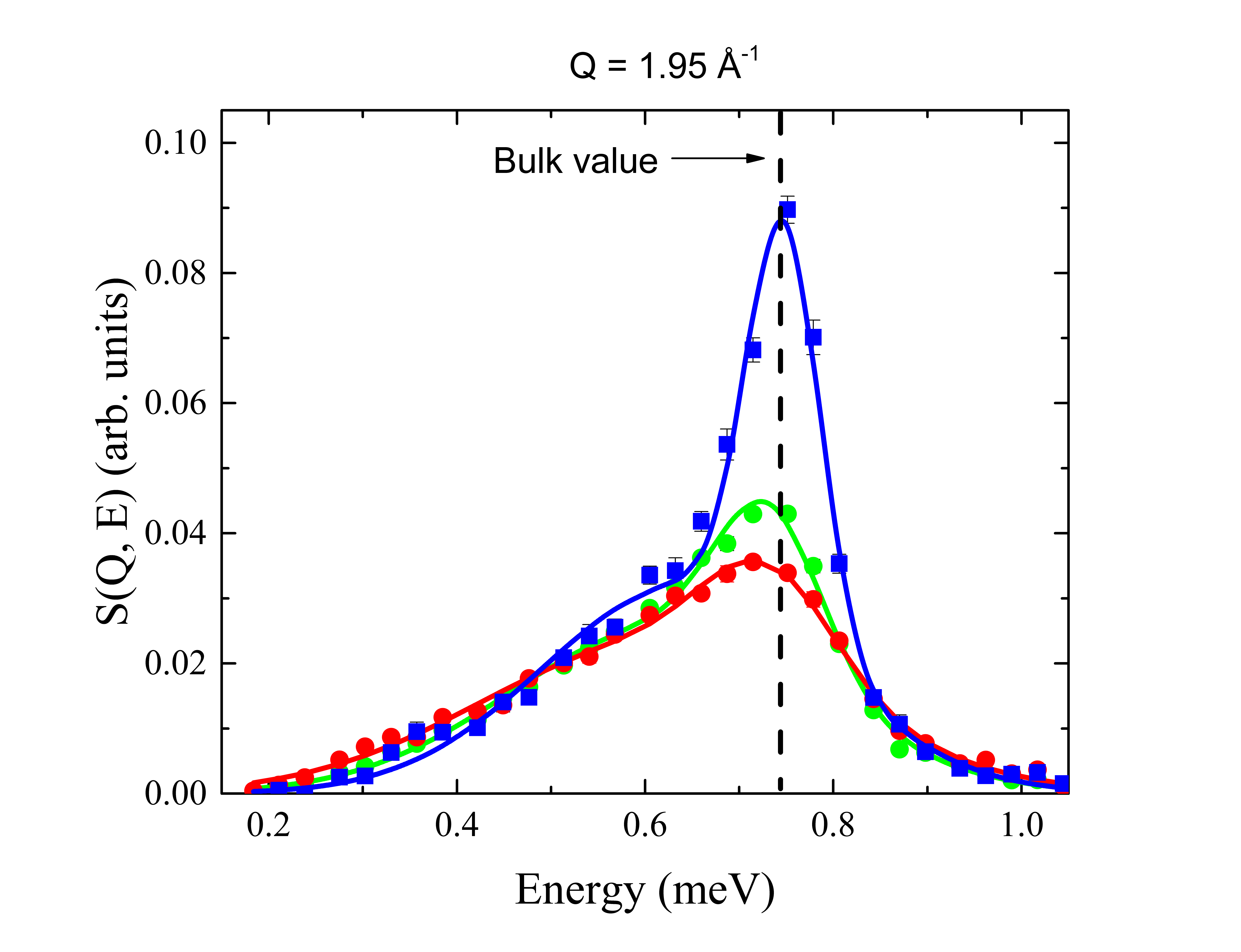}
	\caption{A cut along energy of the roton, Q = 1.95 $\textrm{\AA}^{-1}$. Points and fit lines are correspond to identical fillings as the previous figure. The bulk-like roton and layer mode are summed to make the total fit shown. Two Gaussian components, the bulk-like mode and the layer mode, are required to fit the asymmetric peak in the roton region. The bulk-like scattering is centered near the bulk liquid value, marked with a dashed line. The layer mode is centered at lower energies, is less intense, and broader than the bulk-like mode. As filling increases, the layer mode scattering remains mostly unchanged, while the bulk-like scattering approaches the bulk liquid energy and becomes more intense. }
	\label{fig:RotonFits}
	\end{figure}

Similar constant Q cuts through the scattering at the roton ($Q$=1.95 $\textrm{\AA}^{-1}$) are shown in Fig. \ref{fig:RotonFits}. Again, the ghost roton and background have been subtracted and have yielded a reasonable background subtraction. At the lowest filling the scattering is very broad and asymmetric with a peak below the bulk roton energy.  As the filling increases, a sharp peak near the roton energy emerges while the broad component remains relatively constant.  The broad component has been interpreted as a 2D roton at the pore wall while the sharp component has been interpreted as a more bulk-like liquid in the pore center\cite{dimeo1998localized,plantevin2001excitations,faak2000phonons,thomlinson1980excitations} . At the highest filling studied the bulk-like roton energy approaches that of the bulk while the 2D roton exhibits little filling dependence.  

Detailed values for the excitation energies were obtained from fitting constant $Q$ cuts of the data.  It was convenient to split the data into two regimes.  The first is the maxon region from 0.8 $< Q <$ 1.75 $\textrm{\AA}^{-1}$. The second is the roton region from 1.75 $< Q < $ 2.25 $\textrm{\AA}^{-1}$. In addition to the fits to the excitations, described below, an additional broad, low-intensity Gaussian was used to fit any additional background and multiphonon scattering. This additional background is fixed by scattering far away from the phonon-maxon-roton scattering, and centered above 1.5 meV. The additional background is typically an order of magnitude less intense than the maxon. While there has been some effort to understand the detailed higher energy (E $>$ 1.2 meV) scattering in the bulk liquid\cite{beauvois2016superfluid,gibbs1999collective}, the scattering in confinement appears broad and featureless at energies above the phonon-roton energies.

\begin{figure}
	\includegraphics[width=\linewidth]{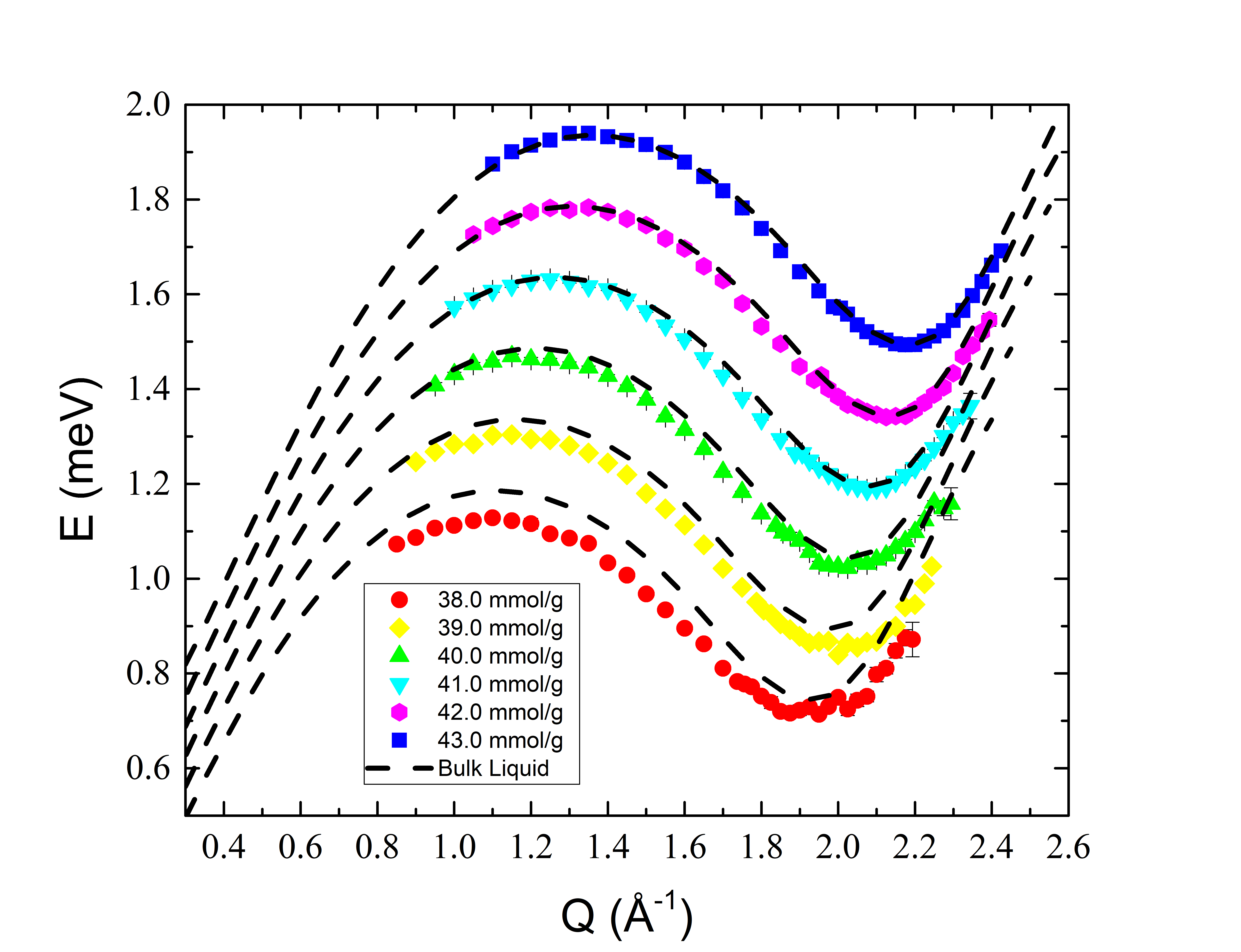}
	\caption{Bulk-like collective excitations as a function of filling. Separate fillings are offset in both energy (0.15 meV each) and momentum transfer (0.05 $\textrm{\AA}^{-1}$ each). The layer mode has been removed for clarity. At each filling, a dashed line indicates the bulk spectrum. }
	\label{fig:Dispersions}
	\end{figure}

The excitations in the maxon region could be fit with a single Gaussian peak. The single Gaussian form gave an excellent fit to the observed scattering, as can be seen in Fig. \ref{fig:MaxonFits}, at all $Q's$ in the maxon region and at all fillings.  The observed peaks were slightly broader than the calculated instrumental resolution indicating there might be some intrinsic lifetime for the maxon. 

\begin{figure}
	\includegraphics[width=\linewidth]{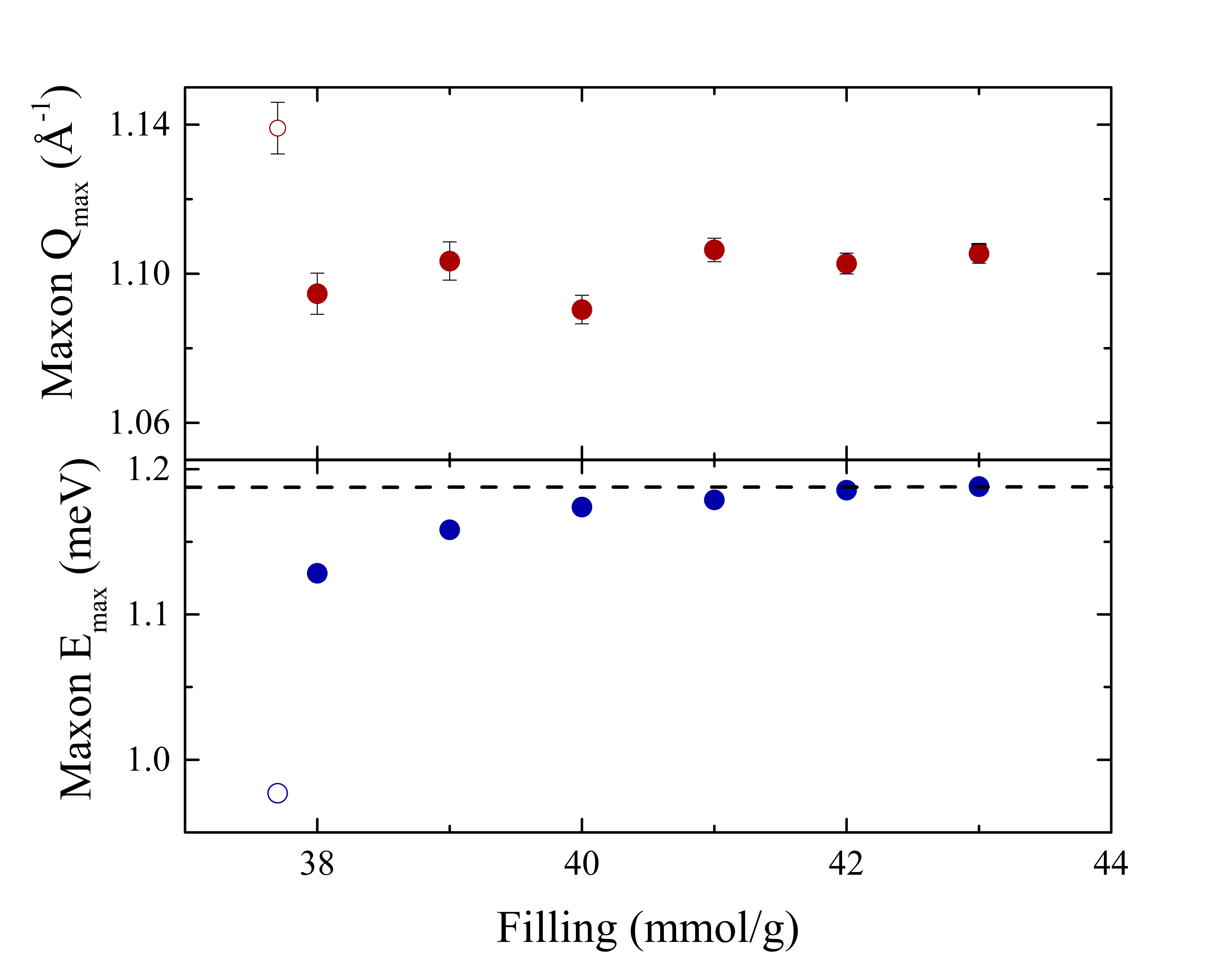}
	\caption{Fitting parameters for the maxon region dispersion as a function of filling. The dashed line indicates the bulk value of the maxon energy. Previous measurements of 37.7 mmol/g\cite{prisk2013phases} are shown as open circles. }
	\label{fig:MaxonParams}
	\end{figure}

The roton region required a sum of two functions to adequately describe the observed scattering as has been used previously\cite{prisk2013phases,albergamo2003elementary,plantevin2002excitations,anderson2002dynamics}.  A narrow Gaussian near the bulk roton energy, the bulk like excitation, and a broader Gaussian well below the bulk, the layer mode, were used.  Previous high resolution measurements of the bulk-like peak using neutron backscattering found the width to be extremely narrow at low temperatures\cite{bryan2017bulklike}, much smaller than our current resolution. Thus, a Gaussian with a width determined by the instrumental resolution was used to fit this component.  The layer mode roton scattering is broad, symmetric, and less intense than the bulk-like mode. This could also be modeled by a Gaussian, but with a much broader width and lower energy.  All fits overlap with the data within error bars. Typical fits are shown in Fig. \ref{fig:RotonFits} where it can be seen that they provide and excellent description of the experimental scattering. However, at low fillings the two peaks are less distinct and the parameters of the two Gaussians are more highly correlated. As filling increases, the low energy component increases slightly in intensity, but not in center or width. The high energy component increases rapidly in intensity and the center changes with filling. At 43.0 mmol/g, the highest filling measured two identifiable, but overlapping peaks appear with the bulk-like component center of 0.743 $\pm$ 0.001 meV in agreement with the bulk value of 0.742 meV\cite{donnelly1981specific,andersen1994collective1,andersen1994collective2}.

The resulting peak centers of the maxon region and the roton region's bulk-like Gaussian are plotted in Fig. \ref{fig:Dispersions} as a function of filling. At the lowest filling (38.0 mmol/g), the entire excitation spectrum is similar to the bulk liquid spectrum, plotted as a black dashed line, but lower in energy across all $Q$. This is in contrast to previous measurements at lower fillings\cite{prisk2013phases}, which measured a reduced maxon energy but an enhanced roton energy. As filling increases, the maxon region of the dispersion monotonically approaches the bulk curve. The roton region at the highest fillings studied is identical to the bulk liquid dispersion. However, as filling increases, the roton dispersion does not monotonically approach the bulk dispersion.

An inverted parabola was used to fit to the measured excitation energies as a function of $Q$ in the maxon region to obtain the maxon energy and $Q_{max}$.  These are plotted as a function of filling in Fig. \ref{fig:MaxonParams}. Previous measurements of 37.7 and 43.0 mmol/g are also plotted for both maxon $Q_{max}$ and energy. The resulting peak energies as a function of filling monotonically approach the bulk value (indicated by a dashed line in Fig. \ref{fig:MaxonParams}) and the previously measured 43.0 mmol/g value. 

\begin{figure}
	\includegraphics[width=\linewidth]{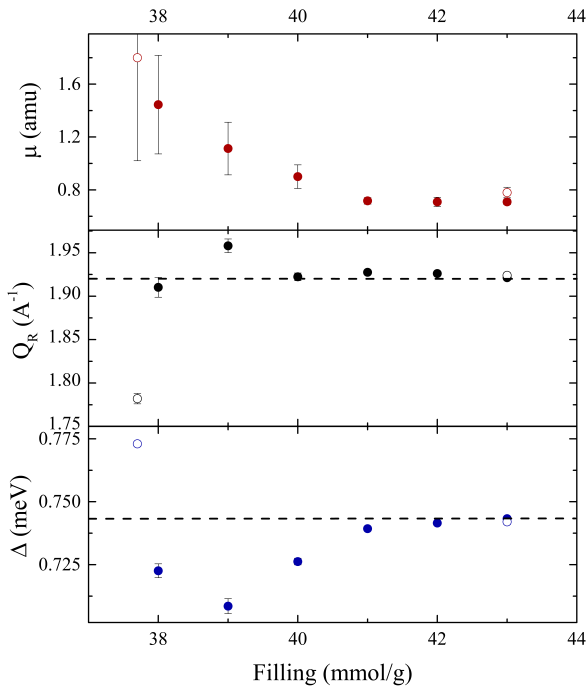}
	\caption{Landau dispersion parameters for the roton as a function of filling. Dotted line indicates bulk value. Previous measurements\cite{prisk2013phases} of 37.7 and 43.0 mmol/g are also shown in pink and blue diamonds, respectively. $Q_{R}$ initially increases until 39.0 mmol/g, then decreases to the bulk value and does not change significantly from 40.0 mmol/g upwards to 43.0 mmol/g. $\Delta$ initially decreases until 39.0 mmol/g, then increases towards the bulk value from below.  }
	\label{fig:RotonParams}
\end{figure}

The bulk-like roton dispersion can be characterized by three parameters, $\Delta$, $Q_{R}$, and $\mu$ which fit the roton dispersion as E $= \Delta$ + $\hbar^2 (Q - Q_{R})^2/2\mu$, known as the Landau dispersion. These parameters are plotted as a function of filling in Fig. \ref{fig:RotonParams}. The effective mass $\mu$ decreases monotonically as a function of filling. Previously measured values of $\mu$ at 37.7 and 43.0 mmol/g indicate a general trend of a monotonically decreasing effective mass, with the current $\mu$ at 43.0 mmol/g in agreement with the previously measured value. A monotonic behavior of $\mu$ including current and previous results is observed. However, this is not the case with $\Delta$ and $Q_{R}$. At the lowest filling studied, $\Delta$ is significantly below the bulk value. As filling is increased, $\Delta$ decreases and then increases towards the bulk value from below. The value of $Q_{R}$ varies within $\pm$ 0.05 $\textrm{\AA}^{-1}$ of the bulk value, with previous thin film measurements indicating a $Q_{R}$ of 1.782 $\textrm{\AA}^{-1}$ which is more than 0.14 $\textrm{\AA}^{-1}$ lower than the bulk value. The current measurements of $\mu$, $\Delta$, and $Q_{R}$ are in agreement with the previously measured values at 43.0 mmol/g.  

Significant scattering below the bulk-like excitation spectrum, the layer mode, is present in this data at all fillings measured. As evident in Fig. \ref{fig:RotonFits}, the layer mode is much broader than the bulk-like mode, and the intensity does not change significantly as a function of filling at $Q_R$. As with other studies of helium confined in porous media, the layer mode dispersion only exists across a limited region in $Q$, near the bulk roton minimum\cite{prisk2013phases,dimeo1998localized,plantevin2001excitations,anderson2002dynamics}. The layer mode does evolve with filling, with a small increase in intensity in the region with energies below the bulk dispersion, at $Q > Q_R$. This change occurs at low fillings, and 40.0 mmol/g and above there is no change in the layer mode scattering. The low filling dispersion is nearly a horizontal line, and evolves into a more parabolic shape at high filling, as well as increasing in energy from 0.49 $\pm$ 0.01 meV at 38.0 mmol/g to 0.62 $\pm$ 0.01 meV at 43.0 mmol/g. However, at low fillings the layer mode peak center is highly correlated with the bulk-like mode peak center. At larger fillings, the correlation is much smaller. Detailed theoretical predictions of the layer mode energy exist, as a function of the substrate well depth and density\cite{apaja2003layer}. Based on those predictions, we estimate a layer mode density of 0.063-0.072 $\textrm{\AA}^{-2}$.

\section{Discussion}

\begin{figure}
	\includegraphics[width=\linewidth]{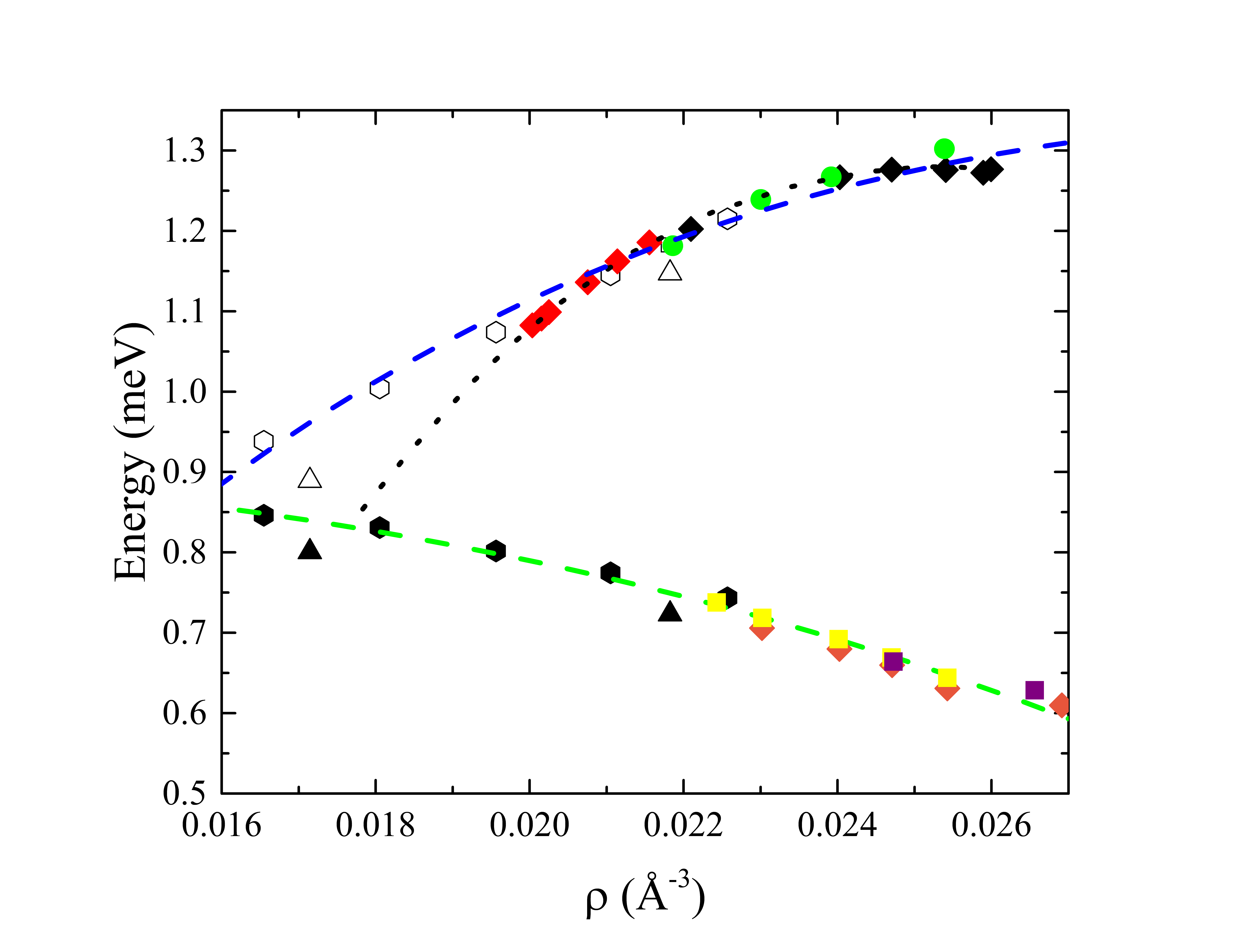}
	\caption{Maxon and roton energy as a function of density. Both theoretical and experimental work is included, with the previous MCM-41 measurements of Albergamo \textit{et. al} (red diamonds), and the cubic fit to the maxon energy (black dotted line) as a function of density also from that work. Bulk liquid maxon energies from previous measurements include neutron scattering data from Andersen (white square)\cite{andersen1994collective1,andersen1994collective2}, Gibbs (green circle)\cite{gibbs1999collective}, and Graf (black diamonds)\cite{graf1974neutron}. DFT results of the maxon (white hexagons) and roton (black hexagons)\cite{maris2002thermodynamic} with Path Integral Monte Carlo maxon energies (white triangles) and roton energies (black triangles)\cite{bauer2000path} are theoretical calculations. Bulk liquid roton measurements are included from Gibbs (yellow squares)\cite{gibbs1999collective}, Dietrich (orange diamonds)\cite{dietrich1972neutron}, and Stirling (purple squares)\cite{stirling1991pressure}. Our fit, as discussed in the text, is shown as a blue dashed line and green dashed line for the maxon and roton, respectively. }
	\label{fig:Context}
	\end{figure}

The density dependence of the excitation spectrum is well known in bulk liquid $^4$He. Several experimental studies have measured the bulk liquid from saturated vapor pressure up to pressures above 20 bar\cite{gibbs1999collective,stirling1991pressure,dietrich1972neutron} . As density increases, the maxon energy increases and the roton energy decreases. Theoretical studies have also examined the bulk liquid below saturated vapor pressure density using Density Functional Theory (DFT)\cite{maris2002thermodynamic} and path integral Monte-Carlo\cite{bauer2000path}. These studies have examined the behavior of the maxon and roton from bulk liquid density (0.0218 $\textrm{\AA}^{-3}$) down to nearly the spinodal point (0.0161 $\textrm{\AA}^{-3}$) where the liquid is no longer stable against separation into a two phase fluid. The theoretical results for densities below the SVP bulk liquid are a smooth continuation of the trends observed at higher densities.  That is, the maxon energy decreases and the roton energy increases with decreasing density. These previous results are shown in Fig. \ref{fig:Context}.

The smooth evolution of the roton and maxon energies with density can be captured using a simple analytical function.  We have used the form $E = E_{0} + B(\rho + \rho_{0})^3$, for excitation energy E and density $\rho$, with constants $E_{0}$, B, and $\rho_{0}$, to fit the behavior of both the maxon and roton energies over the entire density range covered in Fig. \ref{fig:Context}. The DFT results were used for densities below SVP since they agree well with the bulk values and exist at several densities below SVP bulk density.  Equal weight was given to all measurements since the uncertainties of the theoretical points were not known.  The fits do an excellent job of capturing the measured behavior at bulk densities and above as well as the theoretical values at densities below the SVP liquid.  The parameters $E_{0}$, B, and $\rho_{0}$ are listed in Table 1.

\begin{table}[tbp]
\centering
\caption{Parameters of the energy-density relation used in this work for the maxon and roton excitations, with energy in units of meV and density in units of number of atoms per $\textrm{\AA}^{3}$. The function used is of the form $E = E_{0} + B(\rho+\rho_{0})^3$. }
\label{table1}
\begin{tabular}{lcccc}
\hline
      & $E_{0}$        & B    & $\rho_{0}$       &  \\ \hline
Roton & 0.924 & -16800  & 0   &  \\ 
Maxon & 1.35   & 59300 & -0.0358 &  \\ 
\end{tabular}
\end{table}

It has been proposed that the maxon energy could be used to directly measure the density of the core liquid.  Lauter \textit{et al.}\cite{lauter19924he} studied helium films on grafoil and observed maxon energies below the bulk values. They interpreted these as due to a low density liquid resulting from the mismatch with the confining potential of the substrate.  More recently Albergamo \textit{et al.}\cite{albergamo2004phonon} have measured the maxon energy in a small porous system (4.7 nm diameter), MCM-41. They also observed a reduced maxon energy which they attributed to the core liquid being at a negative pressure, stretched apart by the substrate.  A functional form for the density dependence of the maxon in MCM-41 was proposed by Albergamo \textit{et al.}\cite{albergamo2004phonon}, which is also shown in Fig. \ref{fig:Context} for comparison. However, it should be noted that in neither of these studies was the density of the liquid measured independently. 

We have used the maxon-density relation from Albergamo \textit{et al.} to determine the density of the confined liquid in our studies.  These should be qualitatively comparable since the MCM-41 used by Albergamo and the FSM-16 used here both have silica-based hexagonal mesopores with comparable pore sizes. While the maxon energy-density relation of Albergamo, shown in Fig. \ref{fig:Context}, differs slightly from the one we have determined at the lowest densities, they are nearly identical in the region covered by these studies.

\begin{figure}
	\includegraphics[width=\linewidth]{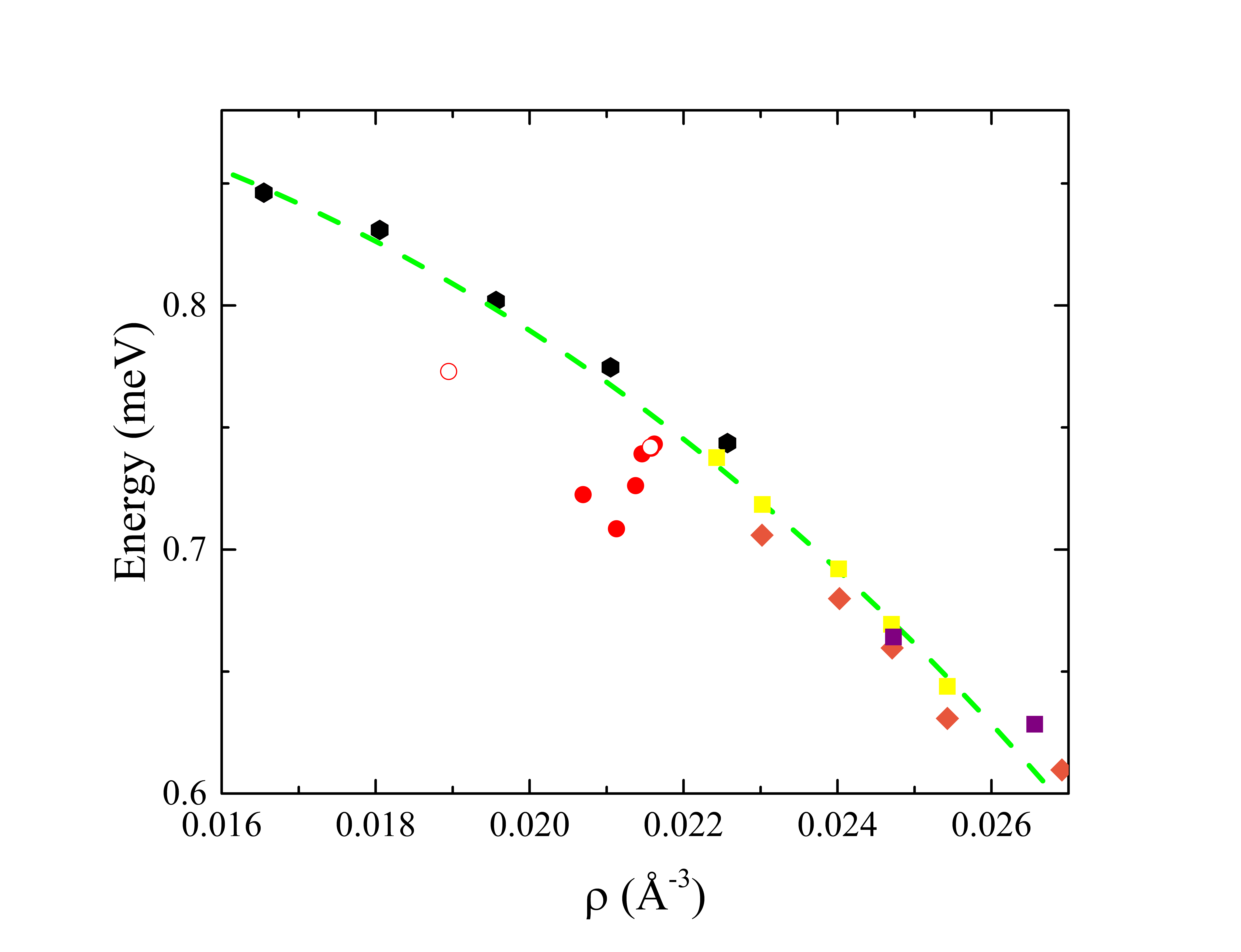}
	\caption{Roton energy plotted as a function of density. Here we have used the observed energy of the maxon and the maxon energy-density curve of Albergamo \textit{et al.} to determine the density for our recent measurements. The black hexagons are from the DFT results of Maris and Edwards\cite{maris2002thermodynamic}. Measurements of $^4$He in confinement include previous FSM-16 measurements (open red circles) \cite{prisk2013phases}, and current measurements (red squares). Bulk liquid roton measurements are included from Gibbs (yellow squares)\cite{gibbs1999collective}, Dietrich (orange diamonds)\cite{dietrich1972neutron}, and Stirling (purple squares)\cite{stirling1991pressure}. The green dashed line is a fit including previous experimental and theoretical work, discussed in the text. The measurements presented here do not scale with density as expected and deviate significantly from the bulk liquid trend. }
	\label{fig:GMDensity}
	\end{figure}

Fig. \ref{fig:GMDensity} shows our measurements of the roton energy as a function of density as determined by the maxon energy. Also shown are the theoretical DFT results as well as several measurements in the bulk liquid. As can be seen, our measured roton energies in confinement do not follow the smooth evolution with density evident in both the previous bulk measurements and the theoretical calculations. Our measurements observe a roton energy that initially decreases with decreasing density, opposite to the expected behavior. As the density, determined by the maxon energy, is further lowered the roton energy begins to increase and approach the theoretically predicted values.  We note that the density here is predicated on the assumption that the confined liquid simply behaves as a bulk liquid at densities below the bulk SVP density, i.e. a liquid under negative pressure. The large discrepancy between the measured and expected roton energies calls into question this assumption.

\begin{figure}
	\includegraphics[width=\linewidth]{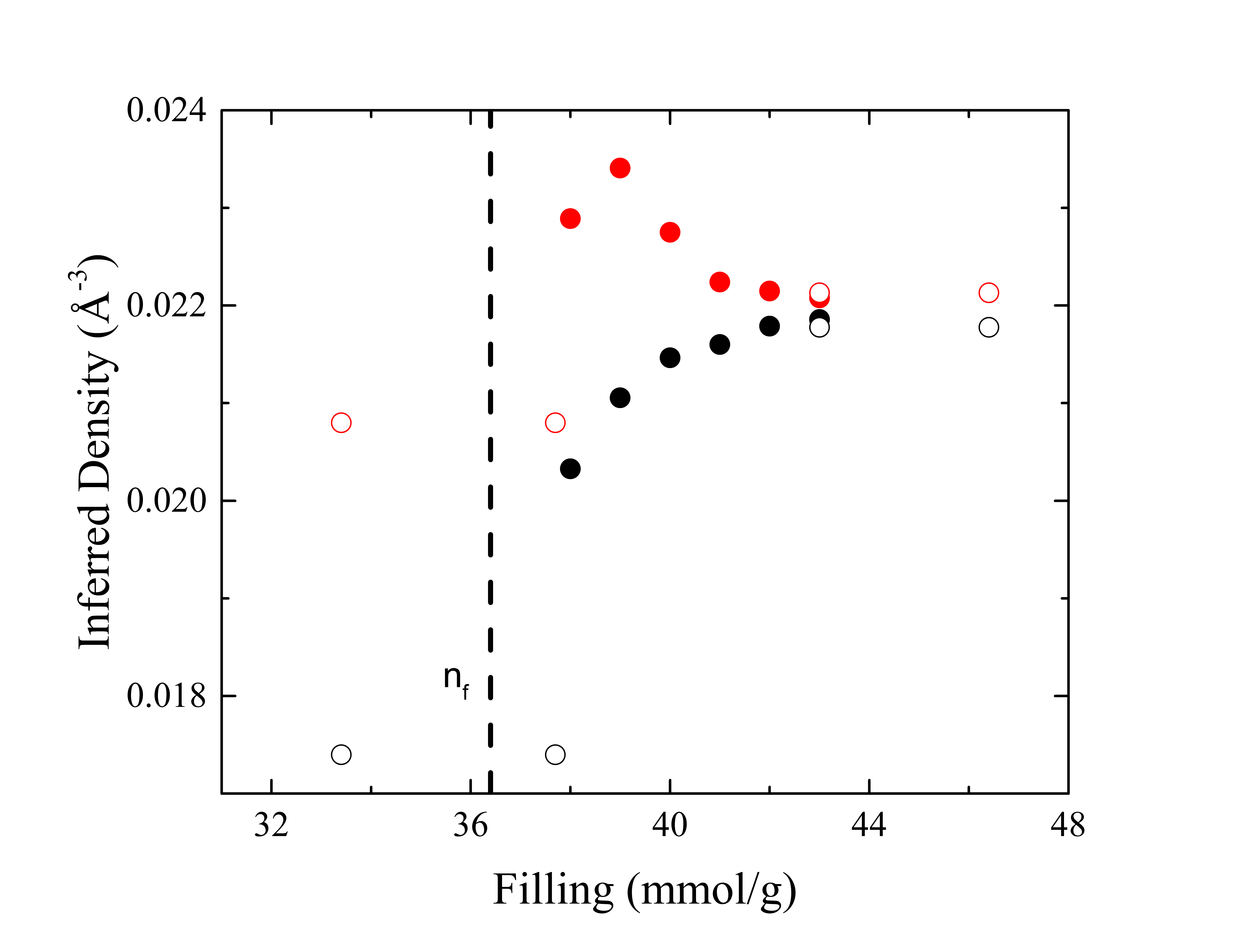}
	\caption{Inferred $^4$He density as a function of filling, for the maxon shown in red and the roton shown in black. The energy-density relationship is shown in the previous figure (Fig. \ref{fig:Context}) and was used to generate inferred density for each maxon and roton energy measured. Approximate first liquid layer completion ($n_{f}$)\cite{wada2008low} is marked with a vertical dashed line. Previous measurements of this sample are included from Prisk\cite{prisk2013phases}, as open black and open red points. }
	\label{fig:InfDensity}
	\end{figure}

This assumption that the confined liquid behaves like the bulk liquid at negative pressure can be examined directly, by using the maxon and roton energies independently to determine an inferred density for the confined liquid.  If the confined liquid behaves as a bulk liquid at negative pressure then the densities inferred from the maxon and roton should agree and follow the same behavior as the pores are filled. The inferred density from both the maxon and roton, using our fits to the bulk values and DFT calculations, are shown in Fig. \ref{fig:InfDensity} as a function of pore filling.  As can be seen, at the highest fillings measured the inferred density from the maxon and roton are similar and approach the bulk density.  However, at lower fillings these inferred densities differ dramatically.  The inferred density from the maxon decreases with decreasing pore filling, as has been observed in previous studies and consistent with the proposal that the confined liquid is a bulk-like liquid at negative pressure.  The roton, however, exhibits a dramatically different behavior.  The inferred density from the roton increases with decreasing pore filling in direct contradiction with both the expected behavior and the density inferred from the maxon energy. At lower pore fillings the inferred density decreases and then remains constant for single layer films.  However, the densities inferred from the maxon and roton never come into agreement at low fillings.

The previous studies of Prisk $\textit{et al.}$ reported that the maxon and roton energies were consistent with values of bulk liquid value for high pore fillings.  They also found that for fillings within the first liquid layer ($21.9 < n < 36.3$ mmol/g) the maxon and roton energies were considerably different from the bulk values and independent of filling. The low filling excitation spectrum is in agreement with calculated spectra of low density, two-dimensional films\cite{apaja2003layer,clements1994dynamics,arrigoni2013excitation}. Present results span the range between the low filling and high filling regions studied by Prisk $\textit{et al.}$ and indicate that there is a non-trivial evolution of the excitations as a function of filling.  We find that the maxon monotonically evolves, with increasing filling, from its value at low filling to the bulk value at high filling.  In contrast, the roton exhibits non-monotonic behavior upon increasing filling.  It evolves from its value at low fillings, which is considerably above the bulk value, to a value much lower than the bulk and then approaches the bulk from below as the filling is further increased.  Examining the behavior of the roton and maxon together clearly illustrates that the confinement in these small pores introduces more subtle effects than simply decreasing the density of a bulk-like liquid. Several previous variational studies\cite{apaja2003excitations,apaja2001layered,apaja2001layered2} have found a reduced roton energy as a result of layering, which may explain the decreasing roton energy from 37.7 to 39.0 mmol/g. However, direct comparisons between theory and experiment of the roton energy as a function of density remain difficult, as the density of the confined liquid in these measurements is unclear.   

At some of the fillings measured here, the excitation spectrum does not resemble a bulk liquid spectrum at any pressure. To fully understand the excitations of the confined liquid, theories that fully account for both the confinement and surface interactions are required.  One possibility is the application of Quantum Monte Carlo methods. New mathematical techniques for extracting $S(Q, E)$ from first-principles are now available\cite{ferre2016dynamic,campbell2015dynamic,dmowski2017observation}, which could be used to understand the confined $^4$He studied here. X-ray Compton scattering may provide another experimental avenue for directly measuring $^4$He density in confinement\cite{tanaka2017cryogenic}. The confined liquid has been studied extensively, across a wide range of temperatures and pressures, and even beyond the roton minimum. However, few measurements exist of the maxon and roton excitations in small pore, partially filled systems, and further measurements could shed light on the complicated relationship between density and the excitation spectrum.  

\section{Conclusions}

Intermediate fillings of superfluid helium confined in FSM-16 have been measured with inelastic neutron scattering, as the thin film transitions into a confined fluid. The scattering of the maxon region indicates a monotonic transition in density from the lower density of the thin film phase to a bulk-like density and spectrum at higher fillings. However, Landau equation parameters and the roton scattering exhibit a more complicated behavior as a function of filling, and do not monotonically increase. The intermediate filling behavior observed in the maxon and roton region does not correspond to a bulk-like dispersion at negative, low, or high pressure. Only at the highest filling measured does the spectrum correspond to a bulk-like dispersion at saturated vapor pressure when full pore filling has been reached.       

\section{Acknowledgements}

This research used resources at the Spallation Neutron Source, a DOE Office of Science User Facility operated by the Oak Ridge National Laboratory.  This report was prepared by Indiana University under award 70NANB10H255 from the National Institute of Standards and Technology, U.S. Department of Commerce.  Matthew S. Bryan acknowledges support under NSF grant DGE-1069091. The authors thank Timothy Prisk for comments and suggestions in preparation of this manuscript.  

\bibliography{CNCStext}

\end{document}